\documentclass{article}
\usepackage{spconf,amsmath,graphicx,multirow,threeparttable}
\usepackage[english]{babel}

\title{Feature reinforcement with word embedding and parsing information in neural TTS}

\name{Huaiping Ming$^1$, Lei He$^1$, Haohan Guo$^2$, Frank K. Soong$^1$}
\address{$^1$Microsoft China\\$^2$School of Computer Science, Northwestern Polytechnical University, Xi'an, China}
\begin{document}
%\ninept
%
\maketitle

\begin{abstract} 
In this paper, we propose a feature reinforcement method under the sequence-to-sequence neural text-to-speech (TTS) synthesis framework. The proposed method utilizes the multiple input encoder to take three levels of text information, i.e., phoneme sequence, pre-trained word embedding, and grammatical structure of sentence from parser as the input feature for the neural TTS system. The added word and sentence level information can be viewed as the feature based pre-training strategy, which clearly enhance the model generalization ability. The proposed method not only improves the system robustness significantly but also improves the synthesized speech to near recording quality in our experiments for out-of-domain text.
\end{abstract}

\section{Introduction} 
\label{sec:intro}

TTS aims at generating a speech waveform from its corresponding text string. Neural TTS systems such as \cite{wang2017tacotron,sotelo2017char2wav,ping2017deep,shen2018natural,ping2018clarinet} have achieved impressive results in recent years. The duration, which is a very important prosody feature for speech, is modeled separately from other prosody and spectral feature for traditional HMM and DNN based TTS pipelines \cite{zen2007hmm,wu2016merlin}. What's more, the duration is usually modeled in very small linguistic units such as in phone level. By contrast, neural TTS systems model all the prosody and spectral feature in sentence level with the help of attention mechanism. The joint optimizing of spectral and prosody feature results in more natural synthesized speech.

There are extensive studies on neural TTS in recent years. Wenfu et al. \cite{wang2016first} took the first step touching neural TTS with attention based sequence-to-sequence model. After that, some neural TTS model such as Tacotron \cite{wang2017tacotron} and Char2Wav \cite{sotelo2017char2wav} were proposed to synthesize speech directly from characters. Tacotron2 \cite{shen2018natural} simplified the network structure of Tacotron, and then uses the WaveNet \cite{van2016wavenet} as a vocoder to synthesize high quality speech close to recordings. Furthermore, there is a preliminary study of semi-supervised training which utilizing textual and acoustic knowledge contained in large unpaired text and speech corpora to improve the data efficiency \cite{chung2018semi}. Some other studies under the framework of attention based neural TTS system including DeepVoice \cite{ping2017deep}, ClariNet \cite{ping2018clarinet}, and VoiceLoop \cite{taigman2018voiceloop}.

The neural TTS system can generate high quality speech, however, there remain challenges too.
The amount of high quality parallel $<$\emph{text, audio}$>$ paired data is quite limited compared with the available data for natural language processing (NLP) tasks such as machine translation \cite{hassan2018achieving}. What's more, the training text is often sourced from very few domains, for instance, conversational text and news. Usually, the training data for TTS cannot cover rich enough text context, and it's common to encounter the out-of-domain problem. Generally, neural TTS systems have difficulties to cope with out-of-domain text, and they may lead to speech with strange prosody and even wrong pronunciation.

%Secondly, the same text can correspond to various different pronunciations, such as different speed, pitch accent, emotion, etc. From a machine learning point of view, TTS is a one-to-many mapping problem. As a result of the one-to-many mapping nature and the small training corpus in restricted domains, it is easy to obtain a biased model. The biased model usually generates less expressive speech with fixed patterns, for example, speech in reading style with flat pitch accent.

Word embedding \cite{wang2015word} and parsing information \cite{dall2016redefining} have been proved to be beneficial in traditional TTS pipelines. In this paper, instead of taking character or phone sequence as the only input, we propose to utilize information from pre-trained word embedding and grammatical structure of sentences to improve the system performance. This can be viewed as feature based pre-training \cite{devlin2018bert}, which borrows knowledge from features generated by models trained with large data corpus. The word embedding is pre-trained with neural machine translation (NMT) task \cite{hassan2018achieving}, which is based on a sequence-to-sequence encoder decoder model with an attention mechanism. The grammatical structure is extracted by the Stanford Parser tool \cite{socher2013parsing,chen2014fast}, which is a statistical parser using knowledge of language gained from hand-parsed sentences.

Both word embedding and grammatical structure are context sensitive features from language related models. The model to generate word embedding and grammatical structure information are trained with very large text data corpus, which means rich text context coverage. Such prior knowledge would help to solve the out-of-domain problem.
%It is well known that speech prosody is largely dependent with word and sentence structure. The model could learn the common pattern of prosody like pause and pitch accent from the prior knowledge contained in word and sentence structure. Therefore, we expect that the added information would help to train a less biased model which generate more natural and expressive speech.

The rest of this paper is organized as follow: In section~\ref{sec:baseline}, we introduce the related work. The details of the proposed feature reinforcement method are described in Section \ref{sec:propsed_method}. In Section \ref{sec:experiment}, the experimental results are presented. Conclusions are drawn in Section \ref{sec:conclusion}.

\begin{figure}[tb]
  \centering
\centerline{\includegraphics[width=8.0cm]{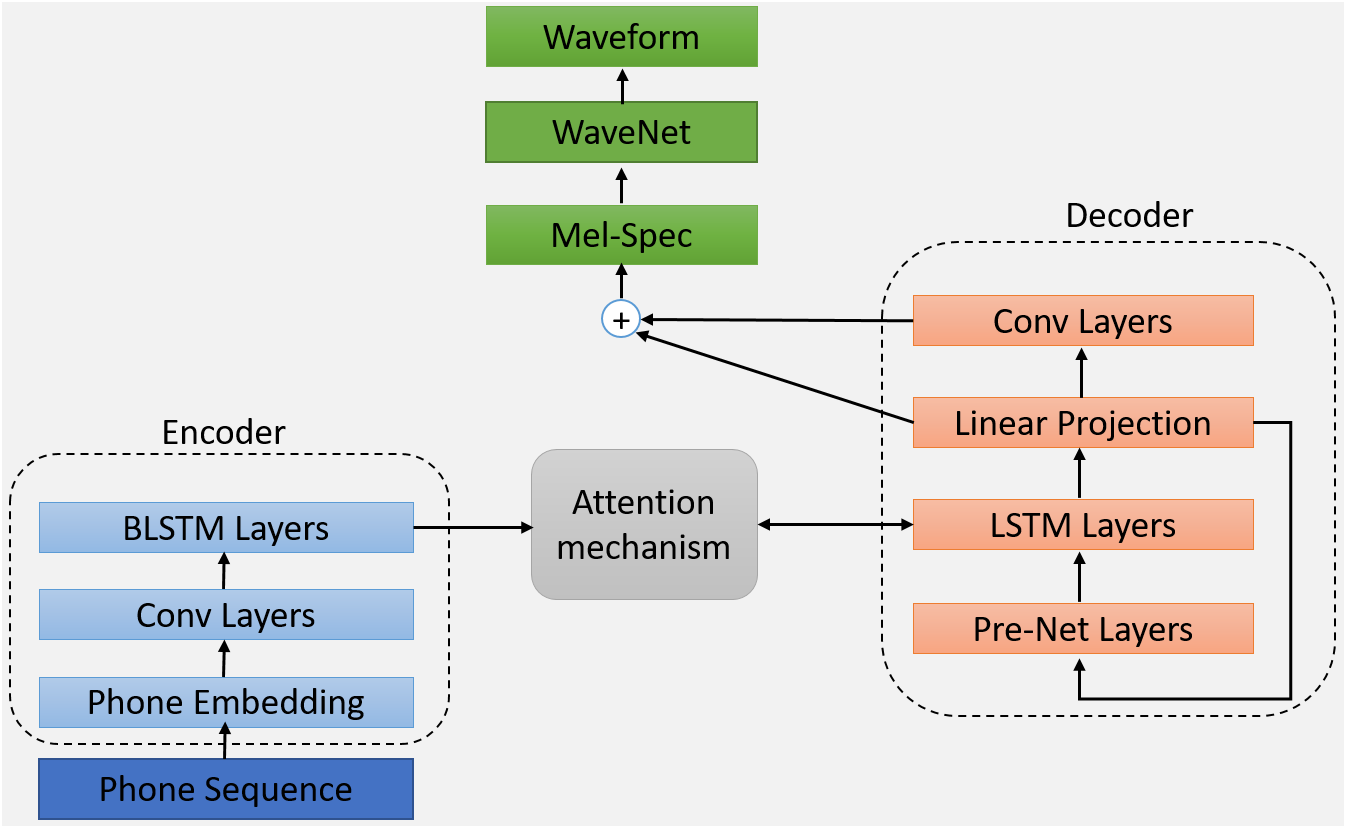}} 
\caption{The baseline network structure of our end-to-end neural TTS system.} \label{fig:baseline} 
\end{figure}

\section{Related Work} 
\label{sec:baseline}
Various linguistic features extracted from text scripts have previously been considered for TTS in traditional HMM and DNN based TTS systems \cite{dall2016redefining}. There are succsessful attempts which utilize word embedding and parsing information as part of the input feature for traditional TTS systems \cite{wang2015word,dall2016redefining}.

Neural TTS represents the state-of-the-art for synthesized speech quality and naturalness.
Currently, state-of-the-art neural TTS \cite{shen2018natural,ping2018clarinet} is based on a sequence-to-sequence encoder decoder model with an attention mechanism.
There are several different network structures that have been proved to be effective for neural TTS system. 

In this paper, we take a re-implemented version of Tacotron 2 \cite{shen2018natural} as the baseline system. Different from the original implementation which takes character sequence as the input, we take phone sequence derived from the normalized text as the input. The network structure of our baseline neural TTS system is illustrated in Fig. \ref{fig:baseline}. In practice,  each phone is mapped to its corresponding random initialized embedding vector. The embedding vectors will be randomly initialized with zero mean and unit variance Gaussian distribution, and they will be jointly updated with the other network parameters during training. The output of the decoder is mel-spectrogram, and it is used as the condition for the WaveNet vocoder to generate corresponding speech waveform.

\section{Feature Reinforcement} 
\label{sec:propsed_method}
\begin{figure*}[t]
\begin{minipage}[b]{0.3\linewidth}
  \centering
  \centerline{\includegraphics[width=6.2cm]{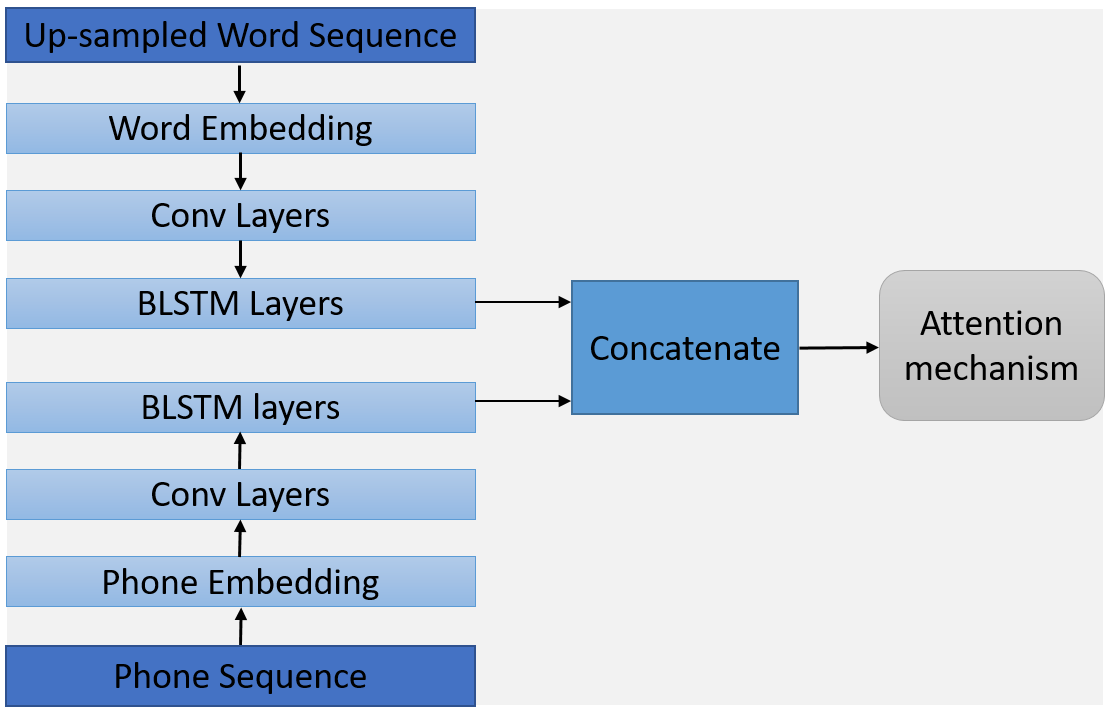}}
  \centering
\end{minipage}\hfill
\begin{minipage}[b]{0.3\linewidth}
  \centering
  \centerline{\includegraphics[width=5.35cm]{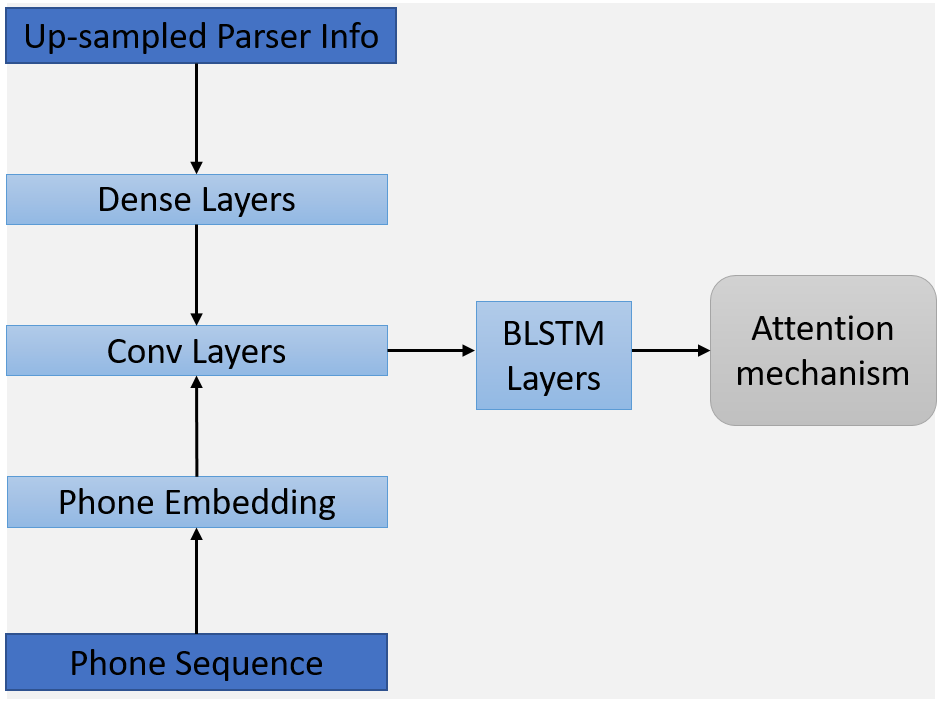}}
  \centering
\end{minipage}
\hfill
\begin{minipage}[b]{0.3\linewidth}
  \centering
  \centerline{\includegraphics[width=6.4cm]{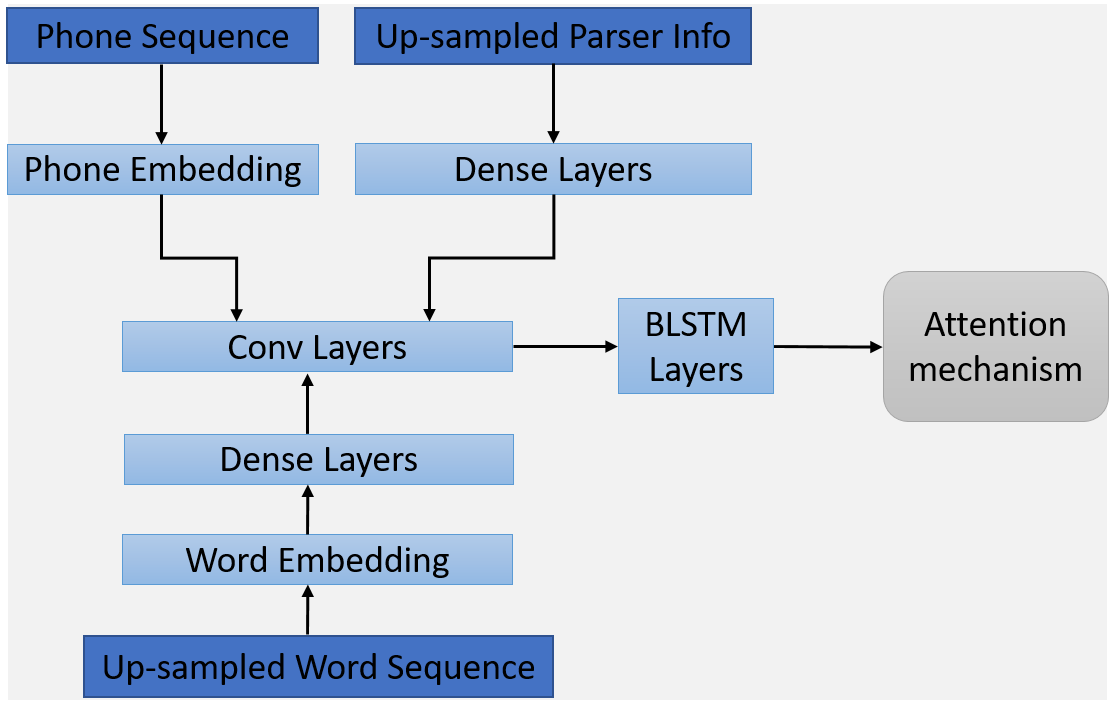}}
  \centering
\end{minipage} 
\caption{The network structure of multiple input encoder for different systems. The left one is the encoder network structure for phone plus word input. The middle on is the encoder network structure for phone plus parser input. The right one is the encoder network structure for phone plus word and parser input. }\label{fig:encoder_structure} 
\end{figure*}

This paper utilizes multiple input encoder to combine information from phone level, word level, and sentence level. The information from different levels of text are mixed together by the encoder and then share the same attention mechanism and decoder. Thus, we only illustrate the network structure of the encoder part for the proposed method as shown in Fig. \ref{fig:encoder_structure}.
With different levels of text information added as the input, we have three different systems. There are many ways of constructing the multiple input encoder. With some comparison experiments, we propose to use different encoder structure for different systems described as follows.

\subsection{Phone plus word input system} 
The word embedding can be obtained from many NLP tasks. Our word embedding is obtained with the system described in \cite{hassan2018achieving}, which achieves human parity on automatic Chinese to English news translation. This system is based on a sequence-to-sequence encoder decoder model with an attention mechanism. The output of the NMT encoder is dumped out as the word embedding. Thus, the word embedding is trained with a similar framework as neural TTS, and we believe this would benefit to the network convergence. The obtained word embedding contains both semantic information and semantic context information. We expect this information would help to solve the out-of-domain problem, and to enrich the prosody of generated speech.

The encoder of the proposed phone plus word input system is illustrated in the left part of Fig. \ref{fig:encoder_structure}.
The word sequence is up-sampled to align with the phone sequence. Specifically, we repeat a word multiple times corresponding to its number of phones. Each word has its corresponding embedding vector pre-trained by NMT task.
The word sequence and phone sequence use separate convolution layers followed by BLSTM layers structure. After that, the output of phone encoder and word encoder are concatenated together and send to the attention mechanism.
 
\subsection{Phone plus grammatical structure input system} 
\begin{figure}[tb]
  \centering
\centerline{\includegraphics[width=9.0cm]{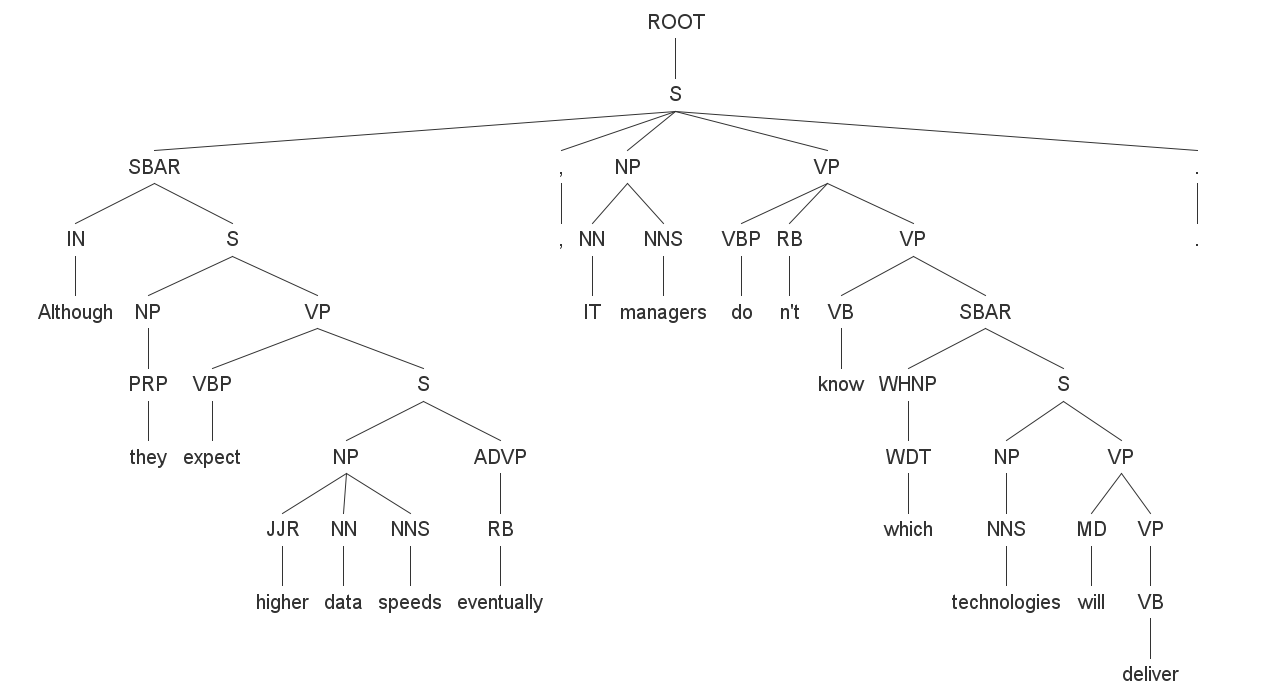}} 
\caption{An example of the grammatical structure of a sentence. The content of this sentence is: ``Although they expect higher data speeds eventually, IT managers don't know which technologies will deliver."} \label{fig:parser} 
\end{figure}

In this system, the grammatical structure of a sentence will be analyzed first of all. We choose the Stanford Parser tool \cite{socher2013parsing,chen2014fast} for grammar parsing. It is a statistical parser using knowledge of language gained from hand-parsed sentences to try to produce the most likely analysis of new sentences. Here we have a parsing example of an English sentence shown in Fig. \ref{fig:parser}. The sentence is parsed into a tree structure according to its grammatical structure. Then we can extract information such as phrase type, phrase border and the relative position of the current word in the current phrase from the parsing results. As the parsing results are extracted in word level, we also need to up-sample the parsing information to align with the phone. The obtained parser features contain from word level to sentence level context-sensitive information. We also expect this information would help to solve the out-of-domain problem, and to improve the prosody performance of generated speech.

The middle part of Fig. \ref{fig:encoder_structure} illustrates the encoder of the phone plus grammatical structure input system. The up-sampled parsing information is passed to dense feed-forward layers to obtain a compressed representation. After that, we concatenate the output of dense layers and the phone embedding, and this mixed information will share the same convolution layers and BLSTM layers to construct the multiple input encoder.

\subsection{Phone plus word and grammatical structure input system}

In this system, we utilize both word embedding and parsing information addition to the phone input. The right part of Fig. \ref{fig:encoder_structure} illustrates the encoder of the phone plus word and grammatical structure input system.  The word embedding and parsing information are obtained exactly the same as the above two systems. Both word embedding and parsing information are passed to their dense layers. Then the output of word and parsing dense layers are concatenated with the phone embedding to mix different level of text information together. After that, the mixed information will share the same convolution layers and BLSTM layers to construct the multiple input encoder.

\section{Experiments} 
\label{sec:experiment}

\subsection{Data Set} 
We tested the proposed systems on a data set recorded by a professional US female speaker. The training data contains approximately 19 hours of speech. The text scripts are in the general domain, and the speech waveform is mono, 16-bit, 16 kHz sampled.
 
\subsection{Naturalness Test}

First of all, a subjective mean option score (MOS) test is conducted for our baseline system. The testing set contains 38 randomly selected in domain texts that are not contained in the training set. Each speech sample is judged by 20 paid native English speakers with a score from 1 to 5 in the test. The MOS test results are shown in Table \ref{tab:mos}. For in domain text, the overall quality of synthesized speech is very close to recordings, and this result is similar to the results reported in \cite{shen2018natural}.

\begin{table}[ht]
\renewcommand{\arraystretch}{1.2}
\caption{The MOS of our baseline system, with 95\% confidence level. In domain text scripts.}\label{tab:mos}
 \centering
\begin{tabular}{|c|c|c|}
\hline
 ~  & Recordings & Baseline \\
\hline
 MOS & 4.51 ($\pm$0.04)  	& 4.41 ($\pm$0.05)   \\
 \hline
\end{tabular} 
\end{table}

Then the overall quality of the proposed method is evaluated by performing MOS tests on 30 out-of-domain texts. The content and style of the selected test text are quite different from the text in the training set, for example, news text which is very long (around 30 words). Each speech sample is also judged by 20 native English speakers with a score from 1 to 5 in this test.

\begin{table}[t]
\renewcommand{\arraystretch}{1.2}
\caption{The MOS of different system and recordings, with 95\% confidence level. Out-of-domain text scripts.}\label{tab:mos_proposed}
 \centering
\begin{tabular}{|c|c|}
\hline
 Systems  & MOS     \\
 \hline
phone input&   4.17 ($\pm$0.06)   \\
\hline
phone + word  input&   4.19 ($\pm$0.06)   \\
\hline
phone + parser input &  4.20 ($\pm$0.06)     \\
\hline
phone + word + parser input &   4.33 ($\pm$0.06)    \\
\hline
Recordings  &  4.44 ($\pm$0.09)    \\
\hline
\end{tabular} 
\end{table}

The test results of different systems are shown in Table \ref{tab:mos_proposed}. The MOS of the baseline system with phone input is 4.17, and the MOS of recordings is 4.44. The quality of synthesized speech has a clear gap to recordings for out-of-domain text. With word embedding onboard, the MOS increased to 4.19. With parser information reinforcement, the MOS increased to 4.20. We can see that adding word embedding and parser information separately will improve the system performance, but the improvement is not significant.

With both word embedding and parser onboard, the MOS increased to 4.33, which is very close to recordings.
The added word and grammatical structure information are complementary, and they together can improve the overall quality of synthesized speech significantly. According to feedback from judges, speech samples synthesized by this system are better in two aspects. Firstly, the pauses between some words sound more appropriate. Secondly, the prosody sound more natural. As expected, the added word and grammatical structure information are beneficial to neural TTS models.

\subsection{Diagnostic Intelligibility Test} 
As mentioned before, neural TTS may generate speech with strange prosody and wrong pronunciation with out-of-domain text. Different from semantically unpredictable sentences (SUS) test, we conduct a diagnostic intelligibility test to evaluate the performance of the proposed system on this problem. First of all, an automatic speech recognition (ASR) tool is utilized to select Griffin-Lim \cite{griffin1984signal} synthesized samples which potentially have pronunciation problems. There are 308 cases selected from around 10000 sentences in different domains. During the test, the judges are requested to mark every unintelligible and unnatural word in the text script. There are two metrics for this test: case level intelligible rate, which is the proportion of the cases without any word marked as ``Unintelligible", and case level natural rate, which is the proportion of the cases without any word marked as ``Unintelligible" or ``Unnatural". Each sentence is marked by one judge, and each judge can marke up to 50 sentences. The judges can listen to the sentences more than once.

The test results for different systems are shown in Table \ref{tab:bad_case}. The case level intelligible rate and natural rate are 88.64\% and 86.36\% respectively for the baseline system with phone input. With word embedding onboard, the intelligible rate and natural rate have 6.49\% and 7.15\% absolute improvement respectively. This means the prior knowledge contained in pre-trained word embedding do help to improve the system robustness for out-of-domain text.

With parser information reinforcement, the system improvement is very clear, which is 7.46\% and 8.77\% for the intelligible rate and natural rate respectively. It's interesting that the improvement is bigger than the results of onboard word embedding. This implies that both word embedding and parser information contains contextual information which improves system robustness, but the parser information contains more useful information.

With both word embedding and parser information, the intelligible rate remains to be 96.10\% compared with phone plus parser input system, but the natural rate increased slightly to 95.45\%. The unchanged intelligible rate implies that most of the contextual information contained in word embedding and parser are nearly equivalent. The increased natural rate further proves that word embedding and parser are complementary for naturalness improvement.

Overall, the proposed system has clear improvements on both intelligible rate (7.46\% absolute improvement) and natural rate (9.09\% absolute improvement) for the selected cases. We can get the conclusion that the added word and grammatical structure information significantly improve the system robustness for out-of-domain text.

\begin{table}[!t]
\newcommand{\tabincell}[2]{\begin{tabular}{@{}#1@{}}#2\end{tabular}}
\renewcommand{\arraystretch}{1.2}
\caption{Diagnostic intelligibility and naturalness test results on 308 selected cases.}\label{tab:bad_case}
 \centering
\begin{tabular}{|c|c|c|}
\hline
 Systems  & \tabincell{c}{Intelligible rate} & \tabincell{c}{Natural rate} \\
 \hline
phone input & 88.64\% & 86.36\% \\
 \hline
phone + word input& 95.13\% & 93.51\%  \\
 \hline
 phone + parser input& 96.10\% & 95.13\%  \\
 \hline
\tabincell{c}{phone + word \\+ parser input}& 96.10\% & 95.45\% \\
 \hline
\end{tabular} 
\end{table}

\section{Conclusion} 
\label{sec:conclusion}

This paper proposes to utilize phoneme level information, word level information, and sentence level information as input features for neural TTS system. The experiments demonstrate that the added features will clearly enhance TTS model generalization ability. The speech naturalness test shows that the proposed method improves the synthesized speech to near recording quality for out-of-domain text. The diagnostic intelligibility test proves that the proposed method significantly improves the system robustness for out-of-domain text. In the future work, we will continue to explore more features from the text, and add them to neural TTS system with different strategies.

\bibliographystyle{IEEEbib}
\bibliography{strings,refs}

\begin{thebibliography}{10}

\bibitem{wang2017tacotron}
Yuxuan Wang, RJ~Skerry-Ryan, Daisy Stanton, Yonghui Wu, Ron~J Weiss, Navdeep
  Jaitly, Zongheng Yang, Ying Xiao, Zhifeng Chen, Samy Bengio, et~al.,
\newblock ``Tacotron: Towards end-to-end speech synthesis,''
\newblock {\em INTERSPEECH}, 2017.

\bibitem{sotelo2017char2wav}
Jose Sotelo, Soroush Mehri, Kundan Kumar, Joao~Felipe Santos, Kyle Kastner,
  Aaron Courville, and Yoshua Bengio,
\newblock ``Char2{W}av: End-to-end speech synthesis,''
\newblock {\em ICLR workshop}, 2017.

\bibitem{ping2017deep}
Wei Ping, Kainan Peng, Andrew Gibiansky, Sercan~O Arik, Ajay Kannan, Sharan
  Narang, Jonathan Raiman, and John Miller,
\newblock ``Deep {V}oice 3: 2000-speaker neural text-to-speech,''
\newblock {\em arXiv preprint arXiv:1710.07654}, 2017.

\bibitem{shen2018natural}
Jonathan Shen, Ruoming Pang, Ron~J Weiss, Mike Schuster, Navdeep Jaitly,
  Zongheng Yang, Zhifeng Chen, Yu~Zhang, Yuxuan Wang, Rj~Skerrv-Ryan, et~al.,
\newblock ``Natural tts synthesis by conditioning wavenet on mel spectrogram
  predictions,''
\newblock in {\em ICASSP}, 2018, pp. 4779--4783.

\bibitem{ping2018clarinet}
Wei Ping, Kainan Peng, and Jitong Chen,
\newblock ``Clari{N}et: Parallel wave generation in end-to-end
  text-to-speech,''
\newblock {\em arXiv preprint arXiv:1807.07281}, 2018.

\bibitem{zen2007hmm}
Heiga Zen, Takashi Nose, Junichi Yamagishi, Shinji Sako, Takashi Masuko, Alan~W
  Black, and Keiichi Tokuda,
\newblock ``The {HMM}-based speech synthesis system ({HTS}) version 2.0.,''
\newblock in {\em SSW}. Citeseer, 2007, pp. 294--299.

\bibitem{wu2016merlin}
Zhizheng Wu, Oliver Watts, and Simon King,
\newblock ``Merlin: An open source neural network speech synthesis system.,''
\newblock in {\em 9th ISCA Speech Synthesis Workshop}, 2016, pp. 202--207.

\bibitem{wang2016first}
Wenfu Wang, Shuang Xu, and Bo~Xu,
\newblock ``First step towards end-to-end parametric {TTS} synthesis:
  Generating spectral parameters with neural attention,''
\newblock in {\em INTERSPEECH}, 2016.

\bibitem{van2016wavenet}
A{\"a}ron Van Den~Oord, Sander Dieleman, Heiga Zen, Karen Simonyan, Oriol
  Vinyals, Alex Graves, Nal Kalchbrenner, Andrew~W Senior, and Koray
  Kavukcuoglu,
\newblock ``Wave{N}et: A generative model for raw audio,''
\newblock in {\em arXiv preprint arXiv:1609.03499}, 2016.

\bibitem{chung2018semi}
Yu-An Chung, Yuxuan Wang, Wei-Ning Hsu, Yu~Zhang, and RJ~Skerry-Ryan,
\newblock ``Semi-supervised training for improving data efficiency in
  end-to-end speech synthesis,''
\newblock {\em INTERSPEECH}, 2018.

\bibitem{taigman2018voiceloop}
Yaniv Taigman, Lior Wolf, Adam Polyak, and Eliya Nachmani,
\newblock ``Voice{L}oop: Voice fitting and synthesis via a phonological loop,''
\newblock {\em arXiv preprint arXiv:1707.06588}, 2018.

\bibitem{hassan2018achieving}
Hany Hassan, Anthony Aue, Chang Chen, Vishal Chowdhary, Jonathan Clark,
  Christian Federmann, Xuedong Huang, Marcin Junczys-Dowmunt, William Lewis,
  Mu~Li, et~al.,
\newblock ``Achieving human parity on automatic chinese to english news
  translation,''
\newblock {\em arXiv preprint arXiv:1803.05567}, 2018.

\bibitem{wang2015word}
Peilu Wang, Yao Qian, Frank~K Soong, Lei He, and Hai Zhao,
\newblock ``Word embedding for recurrent neural network based {TTS}
  synthesis,''
\newblock in {\em IEEE International Conference on Acoustics, Speech and Signal
  Processing (ICASSP)}. IEEE, 2015, pp. 4879--4883.

\bibitem{dall2016redefining}
Rasmus Dall, Kei Hashimoto, Keiichiro Oura, Yoshihiko Nankaku, and Keiichi
  Tokuda,
\newblock ``Redefining the linguistic context feature set for {HMM} and {DNN}
  {TTS} through position and parsing,''
\newblock in {\em INTERSPEECH}, 2016, pp. 2851--2855.

\bibitem{devlin2018bert}
Jacob Devlin, Ming-Wei Chang, Kenton Lee, and Kristina Toutanova,
\newblock ``Bert: Pre-training of deep bidirectional transformers for language
  understanding,''
\newblock {\em arXiv preprint arXiv:1810.04805}, 2018.

\bibitem{socher2013parsing}
Richard Socher, John Bauer, Christopher~D Manning, et~al.,
\newblock ``Parsing with compositional vector grammars,''
\newblock in {\em Proceedings of the 51st Annual Meeting of the Association for
  Computational Linguistics (Volume 1: Long Papers)}, 2013, vol.~1, pp.
  455--465.

\bibitem{chen2014fast}
Danqi Chen and Christopher Manning,
\newblock ``A fast and accurate dependency parser using neural networks,''
\newblock in {\em Proceedings of the conference on empirical methods in natural
  language processing (EMNLP)}, 2014, pp. 740--750.

\bibitem{griffin1984signal}
Daniel Griffin and Jae Lim,
\newblock ``Signal estimation from modified short-time fourier transform,''
\newblock {\em IEEE Transactions on Acoustics, Speech, and Signal Processing},
  vol. 32, no. 2, pp. 236--243, 1984.

\end{thebibliography}

\end{document}